\documentclass[reprint,aps,unsortedaddress,superscriptaddress,prl,floatfix,showpacs]{revtex4-2}
\usepackage[utf8]{inputenc}
\usepackage{graphicx}
\usepackage{amsmath,amsthm,amssymb}
\usepackage{url}
\usepackage{subcaption}
\usepackage[separate-uncertainty=true,range-units=single,range-phrase=\textendash]{siunitx}
\usepackage{physics}
\usepackage{zref-clever}
\zcsetup{cap,abbrev,nameinlink=false}
\usepackage[pdfusetitle]{hyperref}
\hypersetup{pdfauthor={C. H. Leung et al. (FNAL E906/SeaQuest Collaboration)}}
\usepackage{adjustbox}
\usepackage{multirow}
\usepackage{cmap}
\usepackage[version=4]{mhchem}
\usepackage{makecell}
\usepackage{orcidlink}
\usepackage{diagbox}
\graphicspath{{figures/}}
\DeclareSIUnit\barn{b}

\begin{document}

\title{Final SeaQuest results on the flavor asymmetry of the proton light-quark sea with proton-induced Drell-Yan process}
\author{C.~H.~Leung\,\orcidlink{0000-0001-7907-3728}}
\email[Contact author: ]{cleung@jlab.org}
\altaffiliation[Present address: ]{Thomas Jefferson National Accelerator Facility, Newport News, Virginia 23606, USA}
\affiliation{Department of Physics, University of Illinois at Urbana-Champaign, Urbana, Illinois 61801, USA}

\author{J.~Dove}
\affiliation{Department of Physics, University of Illinois at Urbana-Champaign, Urbana, Illinois 61801, USA}

\author{K.~Nagai\,\orcidlink{0000-0002-5336-8306}}
\altaffiliation[Present address: ]{Department of Physics and Materials Science, The University of Memphis, Memphis, TN 38152}
\affiliation{Los Alamos National Laboratory, Los Alamos, New Mexico 87545, USA}
\affiliation{Institute of Physics, Academia Sinica, Taipei, 11529, Taiwan}
\affiliation{Department of Physics, Tokyo Institute of Technology, Meguro-ku, Tokyo 152-8550, Japan}

\author{K.~Nakano\,\orcidlink{0000-0002-8925-2233}}
\affiliation{University of Virginia, Charlottesville, Virginia 22904 USA}
\affiliation{RIKEN Nishina Center for Accelerator-Based Science, Wako, Saitama 351-0198, Japan}

\author{S.~Prasad\,\orcidlink{0000-0003-3404-0062}}
\affiliation{Department of Physics, University of Illinois at Urbana-Champaign, Urbana, Illinois 61801, USA}
\affiliation{Physics Division, Argonne National Laboratory, Lemont, Illinois 60439, USA}

\author{A.~S.~Tadepalli}
\altaffiliation[Present address: ]{Thomas Jefferson National Accelerator Facility, Newport News, Virginia 23606, USA}
\affiliation{Department of Physics and Astronomy, Rutgers, The State University of New Jersey, Piscataway, New Jersey 08854, USA}

\author{C.~A.~Aidala\,\orcidlink{0000-0001-9540-4988}}
\affiliation{Randall Laboratory of Physics, University of Michigan, Ann Arbor, Michigan 48109, USA}
\affiliation{Los Alamos National Laboratory, Los Alamos, New Mexico 87545, USA}

\author{J.~Arrington\,\orcidlink{0000-0002-0702-1328}}
\altaffiliation[Present address: ]{Lawrence Berkeley National Laboratory, Berkeley, California 94720 USA}
\affiliation{Physics Division, Argonne National Laboratory, Lemont, Illinois 60439, USA}

\author{C.~Ayuso}
\affiliation{Randall Laboratory of Physics, University of Michigan, Ann Arbor, Michigan 48109, USA}

\author{C.~L.~Barker}
\affiliation{Department of Engineering and Physics, Abilene Christian University, Abilene, Texas 79699, USA}

\author{W.~C.~Chang\,\orcidlink{0000-0002-1695-7830}}
\affiliation{Institute of Physics, Academia Sinica, Taipei, 11529, Taiwan}

\author{A.~Chen\,\orcidlink{0000-0002-1243-7029}}
\affiliation{Department of Physics, University of Illinois at Urbana-Champaign, Urbana, Illinois 61801, USA}

\author{D.~C.~Christian\,\orcidlink{0000-0003-1275-6510}}
\affiliation{Fermi National Accelerator Laboratory, Batavia, Illinois 60510, USA}

\author{B.~P.~Dannowitz}
\affiliation{Department of Physics, University of Illinois at Urbana-Champaign, Urbana, Illinois 61801, USA}

\author{L.~El~Fassi\,\orcidlink{0000-0003-3647-3136}}
\affiliation{Department of Physics and Astronomy, Mississippi State University, Mississippi State, Mississippi 39762, USA}
\affiliation{Department of Physics and Astronomy, Rutgers, The State University of New Jersey, Piscataway, New Jersey 08854, USA}

\author{D.~F.~Geesaman\,\orcidlink{0000-0003-2557-3131}}
\affiliation{Physics Division, Argonne National Laboratory, Lemont, Illinois 60439, USA}

\author{R.~Gilman\,\orcidlink{0000-0002-7106-2845}}
\affiliation{Department of Physics and Astronomy, Rutgers, The State University of New Jersey, Piscataway, New Jersey 08854, USA}

\author{Y.~Goto\,\orcidlink{0000-0002-2973-7458}}
\affiliation{RIKEN Nishina Center for Accelerator-Based Science, Wako, Saitama 351-0198, Japan}

\author{R.~S.~Guo}
\affiliation{Department of Physics, National Kaohsiung Normal University, Kaohsiung 824, Taiwan}

\author{L.~Guo\,\orcidlink{0009-0002-3562-0552}}
\altaffiliation[Present address: ]{Physics Department, Florida International University, Miami, Florida, 33199, USA}
\affiliation{Los Alamos National Laboratory, Los Alamos, New Mexico 87545, USA}

\author{T.~J.~Hague\,\orcidlink{0000-0003-1288-4045}}
\altaffiliation[Present address: ]{Thomas Jefferson National Accelerator Facility, Newport News, Virginia 23606, USA}
\affiliation{Department of Engineering and Physics, Abilene Christian University, Abilene, Texas 79699, USA}

\author{R.~J.~Holt\,\orcidlink{0000-0001-9225-9914}}
\altaffiliation[Present address: ]{Kellogg Radiation Laboratory, California Institute of Technology, Pasadena, California 91125, USA}
\affiliation{Physics Division, Argonne National Laboratory, Lemont, Illinois 60439, USA}

\author{M.~F.~Hossain\,\orcidlink{0000-0002-6467-1394}}
\affiliation{Department of Physics, New Mexico State University, Las Cruces, NM 88003 USA}
\affiliation{University of Virginia, Charlottesville, Virginia 22904 USA}

\author{L.~D.~Isenhower\,\orcidlink{0000-0002-8237-5636}}
\affiliation{Department of Engineering and Physics, Abilene Christian University, Abilene, Texas 79699, USA}

\author{E.~R.~Kinney\,\orcidlink{0000-0002-4176-5283}}
\affiliation{Department of Physics, University of Colorado, Boulder, Colorado 80309, USA}

\author{A.~Klein}
\affiliation{Los Alamos National Laboratory, Los Alamos, New Mexico 87545, USA}

\author{D.~Kleinjan\,\orcidlink{0000-0002-2737-0859}}
\affiliation{Los Alamos National Laboratory, Los Alamos, New Mexico 87545, USA}

\author{P.-J.~Lin\,\orcidlink{0000-0001-7073-6839}}
\affiliation{Department of Physics, University of Colorado, Boulder, Colorado 80309, USA}
\affiliation{Institute of Physics, Academia Sinica, Taipei, 11529, Taiwan}
\affiliation{Department of Physics, National Central University, Jhongli District, Taoyuan City 32001,Taiwan}

\author{K.~Liu\,\orcidlink{0000-0002-6676-8165}}
\affiliation{Los Alamos National Laboratory, Los Alamos, New Mexico 87545, USA}

\author{M.~X.~Liu}
\affiliation{Los Alamos National Laboratory, Los Alamos, New Mexico 87545, USA}

\author{W.~Lorenzon\,\orcidlink{0000-0003-0657-8463}}
\affiliation{Randall Laboratory of Physics, University of Michigan, Ann Arbor, Michigan 48109, USA}

\author{R.~E.~McClellan\,\orcidlink{0000-0003-0629-4065}}
\altaffiliation[Present address: ]{Pensacola State College, Pensacola, FL 32504, USA}
\affiliation{Department of Physics, University of Illinois at Urbana-Champaign, Urbana, Illinois 61801, USA}

\author{P.~L.~McGaughey}
\affiliation{Los Alamos National Laboratory, Los Alamos, New Mexico 87545, USA}

\author{M.~M.~Medeiros}
\affiliation{Physics Division, Argonne National Laboratory, Lemont, Illinois 60439, USA}

\author{Y.~Miyachi\,\orcidlink{0000-0002-8502-3177}}
\affiliation{Department of Physics, Yamagata University, Yamagata City, Yamagata 990-8560, Japan}

\author{S.~Miyasaka\,\orcidlink{0009-0004-1293-5679}}
\affiliation{Department of Physics, Tokyo Institute of Technology, Meguro-ku, Tokyo 152-8550, Japan}

\author{D.~H.~Morton\,\orcidlink{0000-0003-3813-1375}}
\affiliation{Randall Laboratory of Physics, University of Michigan, Ann Arbor, Michigan 48109, USA}

\author{K.~Nakahara}
\affiliation{Department of Physics, University of Maryland, College Park, Maryland 20742, USA}

\author{J.~C.~Peng\,\orcidlink{0000-0003-4198-9030}}
\affiliation{Department of Physics, University of Illinois at Urbana-Champaign, Urbana, Illinois 61801, USA}

\author{A.~J.~R.~Puckett}
\altaffiliation[Present address: ]{University of Connecticut, Storrs, CT 06269, USA}
\affiliation{Los Alamos National Laboratory, Los Alamos, New Mexico 87545, USA}

\author{A.~Pun}
\affiliation{Department of Physics, New Mexico State University, Las Cruces, NM 88003 USA}

\author{B.~J.~Ramson\,\orcidlink{0000-0002-0925-3405}}
\affiliation{Randall Laboratory of Physics, University of Michigan, Ann Arbor, Michigan 48109, USA}

\author{P.~E.~Reimer\,\orcidlink{0000-0002-0301-2176}}
\affiliation{Physics Division, Argonne National Laboratory, Lemont, Illinois 60439, USA}

\author{J.~G.~Rubin\,\orcidlink{0000-0002-9408-297X}}
\affiliation{Physics Division, Argonne National Laboratory, Lemont, Illinois 60439, USA}
\affiliation{Randall Laboratory of Physics, University of Michigan, Ann Arbor, Michigan 48109, USA}

\author{F.~Sanftl}
\affiliation{Department of Physics, Tokyo Institute of Technology, Meguro-ku, Tokyo 152-8550, Japan}

\author{S.~Sawada\,\orcidlink{0000-0002-7122-1690}}
\affiliation{Institute of Particle and Nuclear Studies, KEK, High Energy Accelerator Research Organization, Tsukuba, Ibaraki 305-0801, Japan}

\author{T.~Sawada\,\orcidlink{0000-0001-5726-7150}}
\affiliation{Randall Laboratory of Physics, University of Michigan, Ann Arbor, Michigan 48109, USA}

\author{M.~B.~C.~Scott\,\orcidlink{0000-0003-1105-1033}}
\altaffiliation[Present address: ]{George Washington University, Washington, DC 20052, USA}
\affiliation{Randall Laboratory of Physics, University of Michigan, Ann Arbor, Michigan 48109, USA}
\affiliation{Physics Division, Argonne National Laboratory, Lemont, Illinois 60439, USA}

\author{T.-A.~Shibata\,\orcidlink{0009-0005-5498-4804}}
\altaffiliation[Present address: ]{Nihon University, College of Science and Technology, Chiyoda-ku, Tokyo 101-8308, Japan}
\affiliation{Department of Physics, Tokyo Institute of Technology, Meguro-ku, Tokyo 152-8550, Japan}
\affiliation{RIKEN Nishina Center for Accelerator-Based Science, Wako, Saitama 351-0198, Japan}

\author{D.-S.~Su\,\orcidlink{0000-0002-8381-7846}}
\affiliation{Institute of Physics, Academia Sinica, Taipei, 11529, Taiwan}

\author{M.~Teo\,\orcidlink{0000-0001-8855-0178}}
\affiliation{Department of Physics, University of Illinois at Urbana-Champaign, Urbana, Illinois 61801, USA}

\author{R.~Towell\,\orcidlink{0000-0003-3640-7008}}
\affiliation{Department of Engineering and Physics, Abilene Christian University, Abilene, Texas 79699, USA}

\author{S.~Uemura\,\orcidlink{0000-0003-3458-4625}}
\altaffiliation[Present address: ]{Fermi National Accelerator Laboratory, Batavia, Illinois 60510, USA}
\affiliation{Los Alamos National Laboratory, Los Alamos, New Mexico 87545, USA}

\author{S.~G.~Wang\,\orcidlink{0000-0001-8474-9817}}
\altaffiliation[Present address: ]{APS, Argonne National Laboratory, Lemont, Illinois 60439, USA}
\affiliation{Institute of Physics, Academia Sinica, Taipei, 11529, Taiwan}
\affiliation{Department of Physics, National Kaohsiung Normal University, Kaohsiung 824, Taiwan}
\affiliation{Fermi National Accelerator Laboratory, Batavia, Illinois 60510, USA}

\author{J.~Wu\,\orcidlink{0000-0003-4432-9521}}
\affiliation{Fermi National Accelerator Laboratory, Batavia, Illinois 60510, USA}

\author{N.~Wuerfel\,\orcidlink{0000-0001-9872-5330}}
\affiliation{Randall Laboratory of Physics, University of Michigan, Ann Arbor, Michigan 48109, USA}

\author{Z.~H.~Ye\,\orcidlink{0000-0002-1873-2344}}
\altaffiliation[Present address: ]{Department of Physics, Tsinghua University, Beijing 100084, China}
\affiliation{Physics Division, Argonne National Laboratory, Lemont, Illinois 60439, USA}

\collaboration{FNAL E906/SeaQuest Collaboration}
\noaffiliation 
\date{\today}

\begin{abstract}
	The Fermilab E906/SeaQuest collaboration performed measurements of the Drell-Yan process using
	\SI{120}{\GeV} proton beams bombarding liquid hydrogen and liquid deuterium targets.
	A combined analysis of all collected data was performed to obtain the final results for the $\sigma_{pd}/2\sigma_{pp}$ Drell-Yan cross
	section ratio covering the kinematic region of $0.13 < x < 0.45$.
	The $x$-dependencies of $\bar{d}\left(x\right) / \bar{u}\left(x\right)$ and $\bar{d}\left(x\right) - \bar{u}\left(x\right)$
	are extracted from these cross section ratios.
	It is found that $\bar{d}\left(x\right)$ is greater than $\bar{u}\left(x\right)$ over the entire measured $x$
	range, with improved statistical accuracy compared to previous measurements.
	The new results on $\bar{d}\left(x\right) / \bar{u}\left(x\right)$ and $\bar{d}\left(x\right) - \bar{u}\left(x\right)$
	are compared to various parton distribution functions and theoretical calculations.
\end{abstract}

\maketitle

Following the discovery of the point-like constituents in the proton in
deep-inelastic scattering (DIS) experiments, evidence for a nucleon sea,
made of quark-antiquark pairs,
was revealed from the observation of the sharp rise of the structure
functions as $x \to 0$. In contrast to the situation for atoms, where
the particle-antiparticle pairs play a relatively minor role, the
quark-antiquark pairs in the nucleon form an integral part for
depicting the internal structure of hadrons, owing to the large
coupling strength $\alpha_s$ in strong interactions~\cite{friedman1972}.

The earliest parton models assumed that the proton sea was SU(2)
flavor symmetric, even though the proton's valence quark
distributions are not flavor symmetric. This flavor symmetry assumption
for the proton sea was not based on any known physics; instead, it reflected the
expectation from perturbative QCD that the splitting of gluons
into quark-antiquark pairs should be nearly up-down flavor symmetric
due to the comparable masses for the up and down quarks.

Evidence for the asymmetry of the $\bar{u}$ and $\bar{d}$ sea-quark
distributions in the proton was first found in
deep-inelastic scattering experiments~\cite{stein1975,amaudruz1991,arneodo1994} via
the observation of the violation
of the Gottfried Sum Rule~\cite{gottfried1967}. Field and Feynman
attributed the asymmetry observed in the early SLAC DIS data to the Pauli
Exclusion Principle~\cite{field1977}.
Many theoretical models were subsequently proposed to explain the
asymmetry, including the meson-cloud model~\cite{thomas1983,speth1997},
the chiral-quark model~\cite{szczurek1996}, and the statistical model~\cite{bourrely2002}.
It was pointed out~\cite{ellis1991}
that an independent experimental technique to probe this flavor asymmetry is
to measure the Drell-Yan cross section ratios,
$\sigma_{pd}/2\sigma_{pp}$, which can further determine the $x$ dependence
of this flavor asymmetry.

The Drell-Yan cross section ratios,
$\sigma_{pd}/2\sigma_{pp}$, have been measured by the CERN NA51 experiment at
\SI{400}{\GeV} for a single value of $x$~\cite{NA51:1994xrz}, and by the FNAL
E866/NuSea experiment for the $0.015 < x < 0.35$ region with an \SI{800}{\GeV}
proton beam~\cite{hawker1998,peng1998,towell2001}. Subsequent measurements by the
HERMES collaboration, using the semi-inclusive DIS reaction~\cite{ackerstaff1998},
and by the STAR collaboration, using $W$-boson production in $p+p$ collisions,
gave further confirmation of the up-down sea-quark flavor asymmetry in
the small $x$ region. The surprisingly large flavor asymmetry of the nucleon
sea has inspired much theoretical work, discussed in several review
articles~\cite{kumano1998,vogt2000a,garvey2001,chang2014,geesaman2019}.

While the small $x$ region of the sea-quark flavor asymmetry was accurately
measured by the FNAL E866/NuSea experiment, the highest $x$ data points from
E866 suggest a possible reversal of the sea-quark flavor asymmetry for $x > 0.25$,
but with large statistical uncertainties. A new measurement with improved
accuracy for the large $x$ region was clearly warranted. The FNAL
E906/SeaQuest experiment, using a new spectrometer~\cite{aidala2019} and
a \SI{120}{\GeV} proton beam, was performed to shed new light on the flavor asymmetry
of the proton at the large $x$ region.

Results from the analysis of the first part of the SeaQuest data, corresponding
to roughly half of the total data collected in the experiment, have been reported
in earlier publications~\cite{dove2021,dove2023}.
These analyses, adopting two different methods, showed that the
$\sigma_{pd}/2\sigma_{pp}$ Drell-Yan cross section ratios remain
above unity, implying that
$\bar{d}$ is greater than $\bar{u}$ for the entire $x$ range measured
at SeaQuest. Results on the $\sigma_{pd}/2\sigma_{pp}$  cross section ratios
for charmonium production, reported recently in
Ref.~\cite{leung2024a}, are also consistent with the flavor asymmetry
deduced from the Drell-Yan data.
Several recent global parton distribution function (PDF)
analyses~\cite{cocuzza2021,ball2022a,accardi2023,alekhin2023,harland-lang2025}
have included these results, in addition to the $W$ boson production data
from the STAR collaboration~\cite{adam2021}, to better constrain the
flavor asymmetry of $\bar u$ and $\bar d$ in the proton. The SeaQuest
flavor asymmetry data have also been compared with the predictions from the
statistical model~\cite{soffer2019}, pion-cloud model~\cite{alberg2022} and the Pauli blocking effect~\cite{chang2022,yin2023}. 

In this paper we present the final results on Drell-Yan $\sigma_{pd}/2\sigma_{pp}$ ratios from the SeaQuest experiment,
including all data collected during the run periods June 2014--July 2015 and March 2016--July 2017.
The combined dataset contains roughly twice as many detected muons as in our previous publications.
Because the trigger conditions and detector configuration differed between the two running periods,
we analyzed each dataset separately.
Results from the two datasets were first compared for consistency and then combined to produce the final results.

The SeaQuest experiment detects $\mu^+\mu^-$ pairs (dimuons) produced in
the interaction of a proton beam with various target nuclei. The production
of massive dimuon pairs is described by the Drell-Yan
process~\cite{drell1970} with the leading order (LO) cross section given as
\begin{multline}
	\frac{d^2\sigma}{dx_1dx_2}=\frac{4\pi \alpha^2}{9x_1x_2s} \times
	\label{eq:DYCross} \\
	\sum_{i\in u,d,s,\dots} e_i^2 \left[q_i^A\left(x_1\right) \bar q_i^B\left(x_2\right) + \bar q_i^A\left(x_1\right)
		q_i^B\left(x_2\right)\right],
\end{multline}
where $\alpha$ is the fine-structure constant,
$s$ is the center-of-mass energy squared,
$e_i$ is the charge of a quark
with flavor $i$, and $q_i^{A,B}\left(x_{1,2}\right)$ are the quark
distribution functions in hadrons $A$ and $B$
for quarks carrying a momentum fraction $x_1$ and $x_2$, respectively.
An analogous notation is used for antiquark distribution functions
$\bar q_i^{A,B}\left(x_{1,2}\right)$.
For fixed-target experiments like SeaQuest, the spectrometers have
large acceptance only for the positive $x_F$ ($x_F = 2p_L/\sqrt{s}\left(1-M^2/s\right)$) region,
where $p_L$ is the longitudinal momentum of the dimuon in the hadron-hadron center-of-mass frame,
and $M$ is the dimuon mass.
Thus, the Drell-Yan cross section is dominated by
the first term, corresponding to the annihilation of a beam quark with
a target antiquark. Assuming
$\sigma^{pd} \approx \sigma^{pp} + \sigma^{pn}$,
which neglects small nuclear effects of the
deuteron~\cite{kumano1998,ehlers2014},
and charge symmetry for the parton distributions~\cite{londergan2010},
\zcref{eq:DYCross} yields the following approximation for the target ratio:
\begin{equation}
	\begin{split}
		\frac{\sigma^{pd}}{2\sigma^{pp}} & \approx
		\frac{1}{2} \frac{4+\frac{d\left(x_1\right)}
			{u\left(x_1\right)}}{4+\frac{d\left(x_1\right)}
			{u\left(x_1\right)}\frac{\bar d\left(x_2\right)}{\bar u\left(x_2\right)}}
		\left[1+\frac{\bar d\left(x_2\right)}{\bar u\left(x_2\right)}\right]                         \\
		                                 & \approx \frac{1}{2} \left[1+\frac{\bar d\left(x_2\right)}
			{\bar u\left(x_2\right)}\right].
	\end{split}
	\label{eq:crRatio}
\end{equation}

While \zcref{eq:crRatio} illustrates the power of the $\sigma^{pd}/2\sigma^{pp}$ Drell-Yan cross section
ratio to reveal the flavor asymmetry between $\bar{d}$ and $\bar{u}$,
the actual extraction of the $\bar{d}\left(x\right) / \bar{u}\left(x\right)$
ratios from the measured $\sigma^{pd}/ 2 \sigma^{pp}$ Drell-Yan cross
section ratios is performed using a Next-to-Leading Order (NLO)
calculation.

The SeaQuest spectrometer is described in detail in Ref.~\cite{aidala2019}.
The experiment receives the proton beam from the Fermilab Main
Injector at \SI{120}{\GeV}
once every minute in 4-second periods (spills), with an average intensity
of \num{6e12} protons per spill.
The target system consists of two liquid targets (hydrogen and deuterium),
three solid targets (iron, carbon, and tungsten), and two calibration
targets (``empty flask'' and ``no target'').
A solid iron magnet focuses the high-mass dimuons into the spectrometer.
The spectrometer consists of four tracking stations, with an open-air magnet between stations~1 and~2.
Each tracking station consists of hodoscope planes, which
provide fast signals for triggering,
and drift chambers or proportional tubes, which provide precise
position information for tracking.
The main physics trigger requires a coincidence of two opposite-sign muons,
one in the top half and one in the bottom half of the spectrometer.
A separate trigger only requires one muon track, and it is mainly used
for studying the accidental background.

The muon tracks are reconstructed using the hits in the drift chambers and proportional tubes in each station.
After identifying hits potentially originating from interactions within the target region,
the Kalman filter algorithm~\cite{kalman1960} is used to reconstruct the muon tracks.
A Kalman filter algorithm~\cite{kalman1960} is used to reconstruct muon trajectories and momenta from hits in the drift chambers and proportional tubes.
Hits originating in the target region are first selected; the Kalman filter is then applied to determine the best-fit tracks and the interaction vertex.
A detailed description of the analysis procedure can be found in earlier publications~\cite{dove2021,dove2023}.

\begin{figure}[htbp!]
	\centering
	\includegraphics[width=\linewidth]{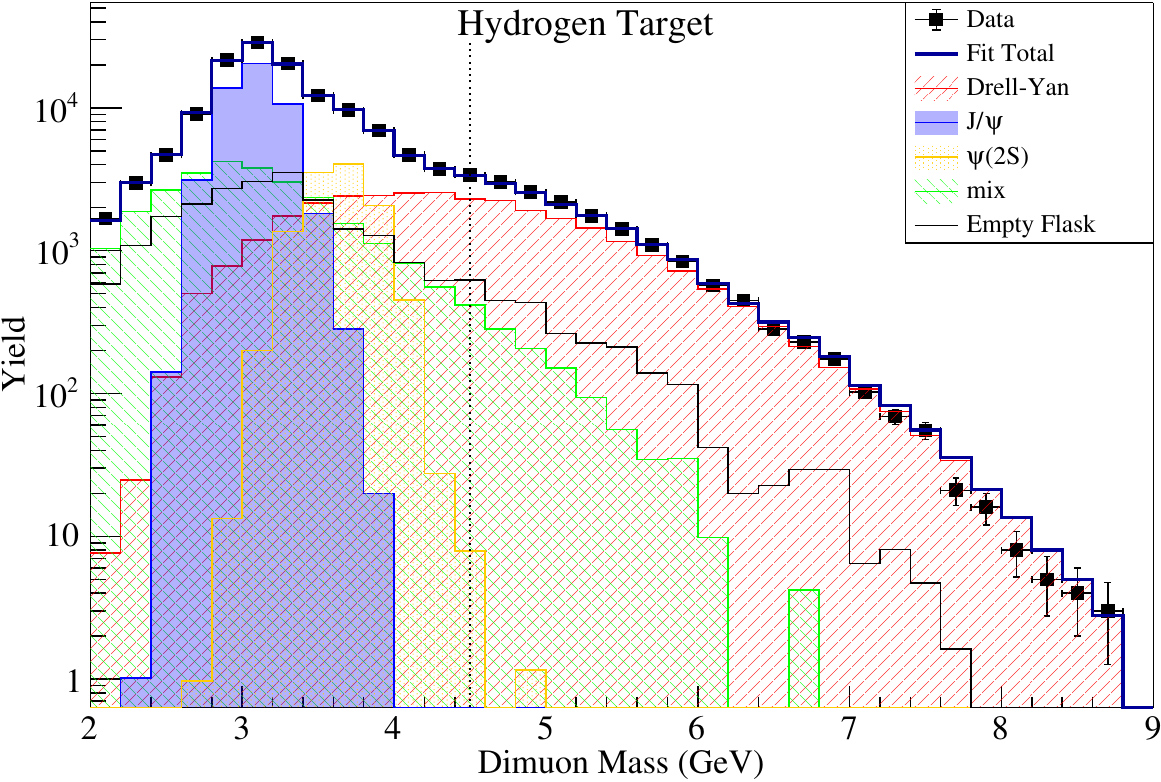}
	\caption{Dimuon mass distribution for events collected
		on liquid hydrogen target for the second data set.
		The data points (solid squares) are compared with a fit consisting of
		various components (see text).}
	\label{fig:massfit}
\end{figure}

A GEANT4~\cite{agostinelli2003,allison2006,allison2016} based Monte
Carlo simulation has been developed to compare data with expectations.
The Drell-Yan events are generated with the dimuon mass and $x_F$
distributions obtained with the CT14 nucleon PDFs~\cite{hou2018}.
The dimuon transverse momentum distribution is adjusted to match the measured distribution~\cite{prasad2020,leung2024}.
The efficiency and the resolution of the chambers are taken into account in the simulation.
An embedding procedure has also been developed to simulate the background hits by embedding each Monte Carlo event
with hits recorded by the random trigger~\cite{dove2020}.

The dimuon mass spectrum, as shown in \zcref{fig:massfit} for data collected
on the liquid hydrogen target for the second part of the experiment, consists
of various components including the Drell-Yan process, the charmonium
production, accidental background, and other background sources.
These different components often have distinct mass spectra, for example, the $J/\psi$
and $\psi\left(2S\right)$ decays would have sharp distributions centered around their masses.
Therefore, the data can be fitted to various templates to obtain
the relative contribution from each source.
As input parameters to the fit, the mass distributions for $J/\psi$, $\psi\left(2S\right)$
and the Drell-Yan are obtained from studying the Monte Carlo simulation. The mass distribution of accidental coincidence events was made by pairing at random $\mu^+$ and $\mu^-$ collected with the ``single-muon'' trigger
under the condition that the beam intensities of $\mu^+$ and $\mu^-$ events were comparable.
The magnitudes of these components,
with the exception of the empty-flask data for which the normalization is known,
were varied in the fit to the mass spectrum.
The statistical uncertainties of the Monte Carlo and empty-flask data are taken into account using the algorithm described in Ref.~\cite{barlow1993}.
\zcref[S]{fig:massfit} shows the fit to the $p+p$ dimuon spectrum,
and the mass distribution is well described by this fitting procedure.

After the mass fit is performed,
a mass cut ($M>\SI{4.5}{\GeV}$) is applied to remove the $J/\psi$ and $\psi\left(2S\right)$ events.
The remaining events are projected into various kinematic variables, such as $x_1$, $x_2$, and $x_F$.
The accidental background and empty-flask contributions are then subtracted from the data.
Several corrections are applied to extract the final Drell-Yan yields.
These include the deadtime correction for the effective luminosity,
and a target contamination correction is also applied to the \ce{D_2} data due to \ce{H} contamination in the target cell.
The reconstruction inefficiency and data acquisition system deadtime, which have a small but non-negligible target dependence,
are also taken into account.
For this analysis,
a revised target contamination correction is used as compared with the earlier work~\cite{dove2021,dove2023},
resulting in a roughly \SI{2}{\percent} upward shift in the measured cross section ratios.

\begin{table*}[htbp!]
	\centering
	\caption{The measured $\sigma_{pd}/2\sigma_{pp}$ cross section ratio as well
		as the extracted $\bar{d}/\bar{u}$ and $\bar{d}-\bar{u}$ for each $x_{2}$ bin.
		The first uncertainty is statistical and the second systematic.
		The average values of kinematic variables in each $x_2$ bin are also shown.}
	\label{tab:dbarubar}
	\begin{adjustbox}{max width=\textwidth}
		{
\renewcommand{\arraystretch}{1.5}
\begin{tabular}{cccccccc}
	\hline\hline
	$x_{2}$ range    & $\expval{x_{2}}$ & $\expval{x_{1}}$ & \begin{tabular}{@{}c@{}} $\expval{p_{T}}$\\(\unit{\GeV/c})\end{tabular} & \begin{tabular}{@{}c@{}}$\expval{M}$\\(\unit{\GeV/c^2})\end{tabular} & $\sigma_{pd}/2\sigma_{pp}$ & $\bar{d}/\bar{u}$                     & $\bar{d}-\bar{u}$                     \\ \hline
	$0.130$--$0.160$ & $0.146$          & $0.687$          & $0.760$                         & $4.71$                        & $1.177\pm0.033\pm0.028$    & $1.383^{+0.058+0.060}_{-0.053-0.060}$ & $0.176^{+0.021+0.024}_{-0.022-0.023}$ \\
	$0.160$--$0.195$ & $0.181$          & $0.610$          & $0.759$                         & $4.87$                        & $1.174\pm0.022\pm0.025$    & $1.431^{+0.041+0.061}_{-0.051-0.061}$ & $0.111^{+0.011+0.011}_{-0.011-0.011}$ \\
	$0.195$--$0.240$ & $0.222$          & $0.553$          & $0.760$                         & $5.11$                        & $1.275\pm0.024\pm0.027$    & $1.672^{+0.052+0.082}_{-0.052-0.082}$ & $0.092^{+0.012+0.012}_{-0.012-0.012}$ \\
	$0.240$--$0.290$ & $0.263$          & $0.516$          & $0.761$                         & $5.44$                        & $1.232\pm0.029\pm0.037$    & $1.653^{+0.073+0.123}_{-0.073-0.113}$ & $0.043^{+0.003+0.013}_{-0.003-0.013}$ \\
	$0.290$--$0.350$ & $0.324$          & $0.492$          & $0.762$                         & $5.83$                        & $1.205\pm0.040\pm0.052$    & $1.694^{+0.124+0.174}_{-0.114-0.174}$ & $0.024^{+0.004+0.004}_{-0.004-0.004}$ \\
	$0.350$--$0.450$ & $0.395$          & $0.474$          & $0.762$                         & $6.34$                        & $1.212\pm0.055\pm0.047$    & $1.925^{+0.205+0.205}_{-0.195-0.205}$ & $0.015^{+0.005+0.005}_{-0.005-0.005}$ \\ \hline\hline
\end{tabular}
}
 	\end{adjustbox}
\end{table*}

\begin{figure}[htbp!]
	\centering
	\includegraphics[width=\linewidth]{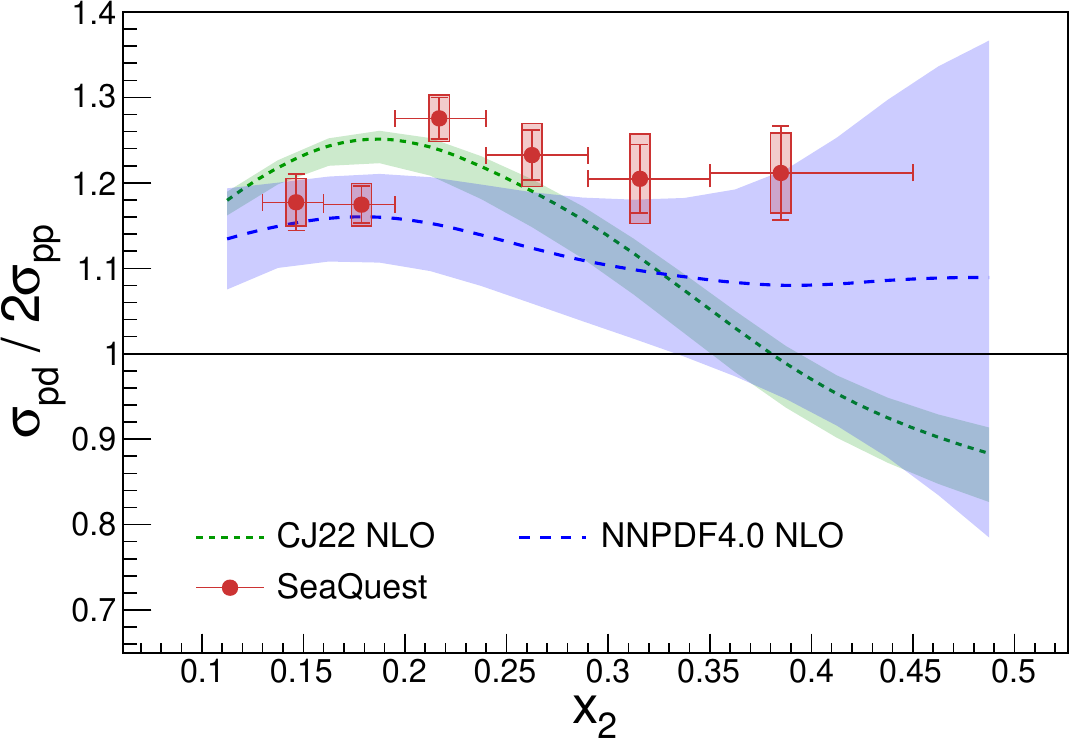}
	\caption{Measured $\sigma_{pd}/2\sigma_{pp}$ Drell-Yan cross section ratio from SeaQuest compared calculations using
		CJ22~\cite{accardi2023}, and NNPDF4.0~\cite{ball2022a}.
		The error bands on the theoretical calculations correspond to the \SI{68}{\percent} confidence level from the PDFs.}
	\label{fig:xT_csr}
\end{figure}

The $\sigma_{pd}/2\sigma_{pp}$ cross
section ratios versus $x_2$ obtained from the analysis of the full SeaQuest
data set are shown in \zcref{fig:xT_csr}.
The ratios as functions of other variables are shown in \zcref{fig:csr_x1_xF} in the End Matter.
The main systematic uncertainties originate from modeling the
accidental background (\qtyrange{0.41}{3.82}{\percent}).
Different methods for generating the accidental background distributions
have been studied~\cite{pate2023} to estimate the uncertainty,
and the differences are quoted as the systematic uncertainty.
Other sources of systematic uncertainties include the efficiency
corrections (\qtyrange{0.46}{0.74}{\percent}),
empty flask normalization (\qtyrange{0.10}{0.16}{\percent}), and the
relative beam luminosity (\SI{2}{\percent}).

The two data sets are analyzed separately, and compared for consistency.
The final results are obtained by taking an average,
weighted by the inverse of the total uncertainty squared.
Except for the smallest $x_2$ point, the results from the two data sets are in good agreement,
with differences less than $1\sigma$.
The flavor asymmetry is very well constrained for the smallest $x_2$ bin by previous experiments,
and the final result on the measured cross section at the smallest $x_2$ bin is in better agreement
with theoretical expectations than the individual data sets.
The results are shown in \zcref{fig:xT_csr} and tabulated in \zcref{tab:dbarubar}.
\zcref[S]{fig:xT_csr} compares the SeaQuest results with
NLO calculations using DYTurbo~\cite{camarda2020},
with nucleon PDFs  CJ22~\cite{accardi2023}, and NNPDF4.0~\cite{ball2022a}.
Both PDFs are obtained from global analyses including
the earlier SeaQuest results~\cite{dove2021}.
This partly explains the better agreement between the data and the NNPDF4.0 calculation, although the data tend to lie above the calculation at large $x$.
Nevertheless, \zcref{fig:xT_csr}
shows that both the data and the NNPDF4.0 calculation for the
$\sigma_{pd}/2\sigma_{pp}$ cross section ratio exceed unity over the
entire measured $x$ region.
With the increased statistics in the updated analysis,
the constraint on the light sea-quark asymmetry in future global analyses can be further improved.

To facilitate comparison between the SeaQuest result and predictions
from various models, the $\bar{d}(x)/\bar{u}(x)$ ratio is extracted from the
cross section ratio using an iterative method described in
Ref.~\cite{dove2021}.
We first estimate the $\bar{d}(x)/\bar{u}(x)$ ratio over the measured $x_2$
and calculate the cross section ratio $R$ using a chosen PDF set.
This PDF set provides all parton distributions including the
$\bar{d}(x)+\bar{u}(x)$, except that the $\bar{d}(x)/\bar{u}(x)$
ratio is allowed to vary.
The calculated cross section is weighted by the spectrometer acceptance, determined using Monte Carlo simulations.
The acceptance is tabulated in \zcref{tab:acceptance}.
The cross section ratios $R$ are then computed at next-to-leading order
as functions of $x_2$ and compared with the data. The $\bar{d}(x)/\bar{u}(x)$
ratios are adjusted until the difference between data and
calculation is less than $10^{-3}$.

\begin{figure*}[htpb!]
	\centering
	\begin{subfigure}{0.49\linewidth}
		\includegraphics[width=\linewidth]{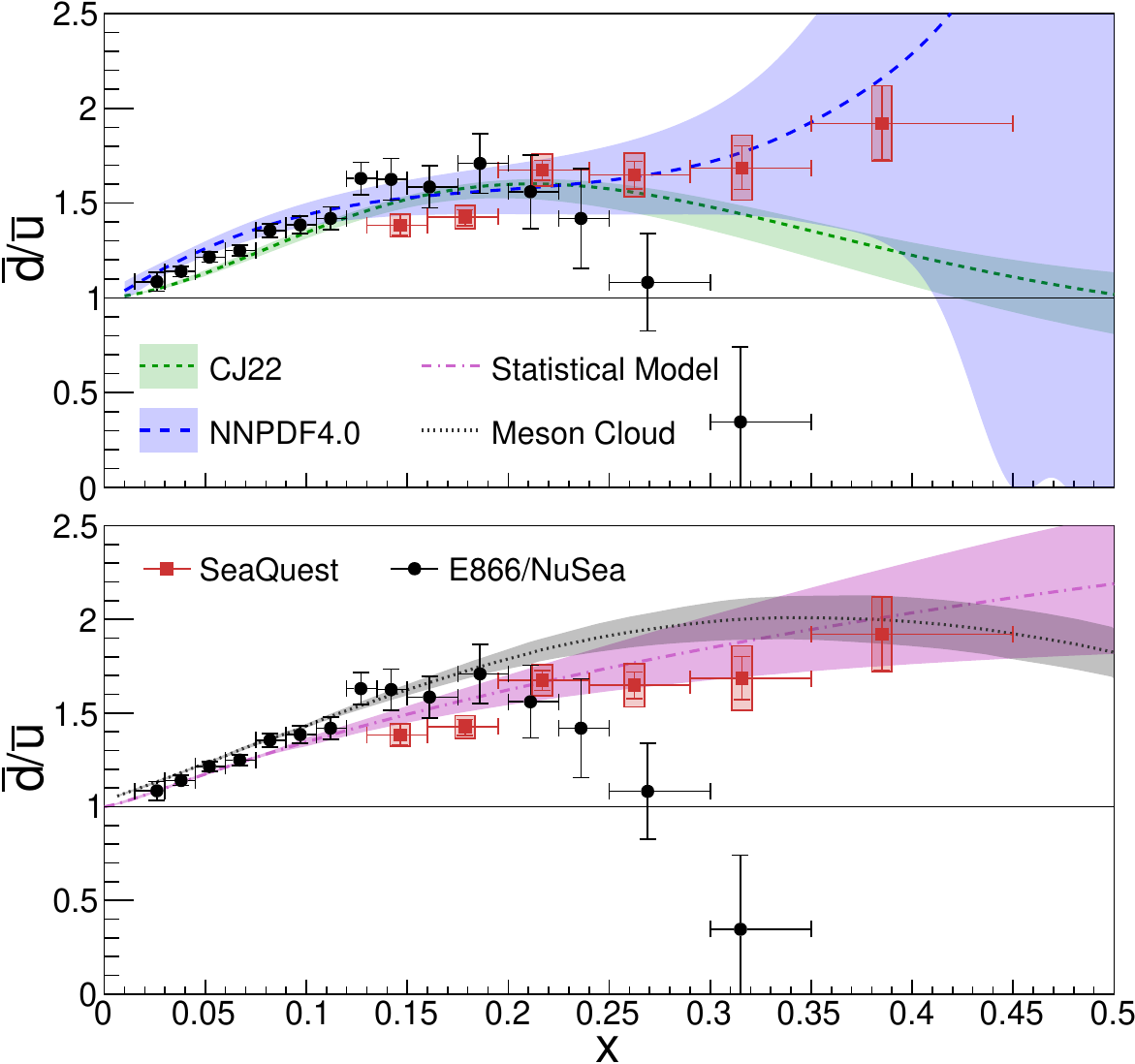}
	\end{subfigure}
	\begin{subfigure}{0.49\linewidth}
		\includegraphics[width=\linewidth]{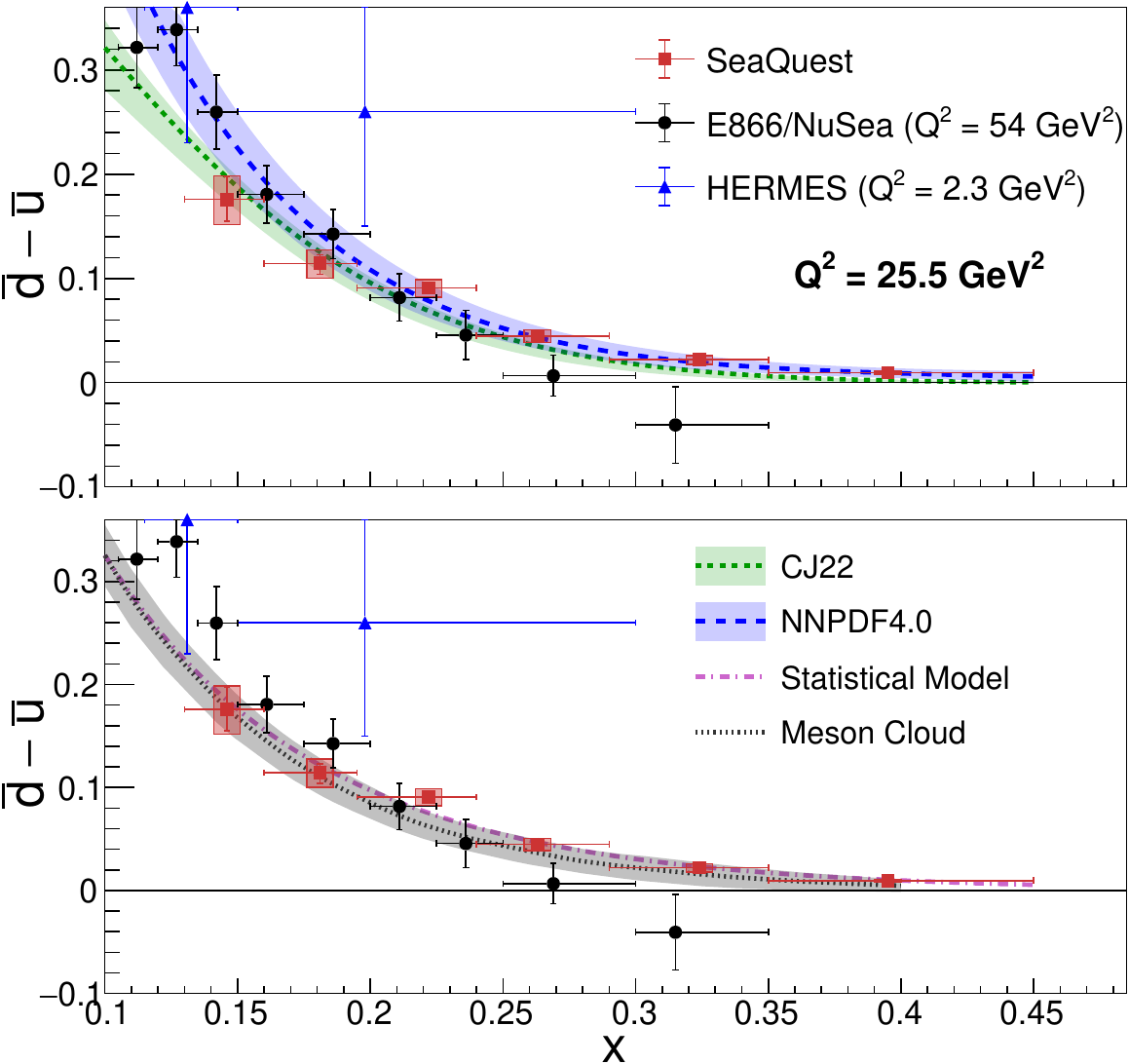}
	\end{subfigure}
	\caption{The extracted $\bar{d}(x)/\bar{u}(x)$ (left) and $\bar{d}(x)-\bar{u}(x)$ (right)
		from the measured $\sigma_{pd}/2\sigma_{pp}$ Drell-Yan cross section ratio
		from SeaQuest (red squares), E866~\cite{towell2001} (black circles), and HERMES~\cite{ackerstaff1998} (blue triangles).
		The extracted ratios are compared with CJ22~\cite{accardi2023}, and NNPDF4.0~\cite{ball2022a} global PDF analyses (top),
		and with predictions from statistical model~\cite{soffer2019} and meson cloud model~\cite{alberg2022} (bottom).
		The error bands on the PDFs correspond to their \SI{68}{\percent} confidence level.
	}
	\label{fig:e906_e866_dbarubar}
\end{figure*}

The extracted $\bar{d}(x)/\bar{u}(x)$ from the measured SeaQuest cross section ratio,
using the CT18 PDFs as the basis, is shown in \zcref{fig:e906_e866_dbarubar}
and tabulated in \zcref{tab:dbarubar}.
While the SeaQuest results are consistent with previous E866
results~\cite{towell2001} at low $x$,
the extracted $\bar{d}(x)/\bar{u}(x)$ ratios from the SeaQuest measurement
continue to rise as $x$ increases,
and are in tension with the E866 result.
The extracted $\bar{d}(x)/\bar{u}(x)$ ratios are compared with results
from CJ22~\cite{accardi2023} and NNPDF4.0~\cite{ball2022a}
global analyses in \zcref{fig:e906_e866_dbarubar}.
NNPDF4.0 and the CJ22 global analyses, which both include the earlier SeaQuest results~\cite{dove2021},
both prefer $\bar{d}(x)/\bar{u}(x)>1$ in the large $x$ region covered by the SeaQuest measurement.
With the improved statistical accuracy from the combined analysis
presented in this paper, the uncertainty on $\bar{d}(x)/\bar{u}(x)$ in
these global analyses can be further reduced.
The SeaQuest results are also in much better agreement with
predictions from the meson cloud model~\cite{alberg2022}
and the statistical model~\cite{soffer2019}.

\begin{table*}[htbp!]
	\centering
	\caption{Values of $\int_{0.13}^{0.45} \left[\bar{d}\left(x\right) - \bar{u}\left(x\right)\right] \dd{x}$
		and $\int_{0.13}^{0.45} x\left[\bar{d}\left(x\right) - \bar{u}\left(x\right)\right] \dd{x}$ at $Q^2=\SI{25.5}{\GeV\squared}$ extracted from
		SeaQuest compared with  CJ22~\cite{accardi2023}, NNPDF4.0~\cite{ball2022a}
		PDFs as well as the statistical models~\cite{soffer2019} and the meson cloud~\cite{alberg2022}.}
	\label{tab:dbarMubar}
	\begin{adjustbox}{max width=\linewidth}
		\renewcommand{\arraystretch}{1.5}
\begin{tabular}{ccccccc}
\hline \hline
 &          & \multicolumn{2}{c}{PDFs} & & \multicolumn{2}{c}{Models} \\ \cline{3-4} \cline{6-7}
 & SeaQuest &CJ22     & NNPDF4.0     & & Stat.     & Meson cloud    \\ \hline
$\int^{0.45}_{0.13} \left[\bar{d}\left(x\right) - \bar{u}\left(x\right) \right]\dd{x}$ &
  $0.0179_{-0.0018}^{+0.0017} {}_{-0.0023}^{+0.0022}$ &
  $0.0167^{+0.0009}_{-0.0028}$ &
  $0.0208^{+0.0036}_{-0.0036}$ & &
  $0.0186$ &
  $0.0180$ \\
$\int^{0.45}_{0.13} x\left[\bar{d}\left(x\right) - \bar{u}\left(x\right) \right]\dd{x}$ &
  $0.00368_{-0.00036}^{+0.00034} {}_{-0.00049}^{+0.00045}$ &
  $0.00319^{+0.00019}_{-0.00063}$ &
  $0.00414^{+0.00078}_{-0.00078}$ & &
  $0.00386$ &
  $0.00361$ \\ \hline \hline
\end{tabular}
 	\end{adjustbox}
\end{table*}

The isovector quantity $\bar{d}(x) - \bar{u}(x)$,
is of interest since perturbative processes should not produce any significant difference between $\bar{d}(x)$ and $\bar{u}(x)$.
Hence, $\bar{d}(x) - \bar{u}(x)$ can provide a direct measure of the non-perturbative contribution to the proton sea.
From the $\bar{d}(x) / \bar{u}(x)$ ratios extracted from SeaQuest,
we have calculated $\bar{d}(x) - \bar u(x)$ by taking the $\bar{d}(x) + \bar{u}(x)$ values from the CT18 proton PDFs.
The values of $\bar{d}(x)-\bar{u}(x)$ derived from the SeaQuest data are
shown in \zcref{fig:e906_e866_dbarubar} and tabulated in \zcref{tab:dbarubar},
and are compared with results from E866~\cite{towell2001} and
HERMES~\cite{ackerstaff1998}.
The derived $\bar{d}(x)-\bar{u}(x)$ are also compared with predictions from the meson cloud and the statistical models in \zcref{fig:e906_e866_dbarubar}.
Like the $\bar{d}(x) / \bar{u}(x)$ ratio, the deduced $\bar{d}-\bar{u}$ is also in good agreement with both models.

The SeaQuest data can also be used to calculate two integrals:
$\int^{0.45}_{0.13} \left[\bar{d}\left(x\right) - \bar{u}\left(x\right) \right]\dd{x}$,
which describe the integrated sea-quark flavor asymmetry, and
$\int^{0.45}_{0.13} x\left[\bar{d}\left(x\right) - \bar{u}\left(x\right) \right]\dd{x}$,
which represents the difference in the momentum fractions carried
by $\bar{u}$ and $\bar{d}$ quarks.
These values are listed in \zcref{tab:dbarMubar}, and compared with different PDFs and models.
The SeaQuest results are in good agreement with NNPDF4.0 and with the model predictions.

In summary, we have reported the improved results on the
$\sigma_{pd}/2\sigma_{pp}$ Drell-Yan cross section ratio up to $x_2=0.45$,
based on the analysis of the entire SeaQuest dataset,
which supersedes the earlier results in Ref.~\cite{dove2021,dove2023}.
The measured ratio remains above unity,
which suggests the $\bar{d}(x)/\bar{u}(x)$ ratio continues to increase at large $x$.
The improved statistical precision presented in this paper will help to
further constrain the light sea-quark asymmetry in future global analyses.
These results support the excess of $\bar{d}$ compared to $\bar{u}$ that is obtained in various model predictions,
including the meson cloud model and the statistical model.
With recent developments in lattice calculations within the framework
of large momentum effective theory~\cite{constantinou2021},
these results can provide an important benchmark for future lattice
calculations.
With the advent of the Electron Ion Collider, the flavor asymmetry in the small $x$ region,
where abundant sea quarks reside, can be further measured via the semi-inclusive DIS reaction~\cite{ackerstaff1998}.

\begin{acknowledgments}
	We thank the late G.~T.~Garvey for contributions to the early stages of this experiment, C.~N.~Brown for contributions to 5 decades of dimuon experiments at Fermilab, and N.~C.~R.~Makins for contributions to the execution of the experiment. We also thank the Fermilab Accelerator Division and Particle Physics Division for their support of this experiment. This work was performed by the SeaQuest Collaboration, whose work was supported in part by the
	U.S. Department of Energy, Office of Nuclear Physics under contract
	No.\ DE-AC02-06CH11357, Grants No.\ DE-FG02-03ER41243, and No.\ DE-FG02-07ER41528; the US National Science Foundation under Grants No.\
	PHY\ 2013002, PHY\ 2110229, PHY\ 2111046, PHY\ 2209348, PHY\ 2309922, and PHY\ 2514181; the JSPS (Japan) KAKENHI through Grants No.\ 21244028, No.\ 25247037, No.\ 25800133, No.\ 18H03694, No.\ 20K04000, No.\ 22H01244, No.\ 23K22515, and No.\ 25K07341; and the National Science and Technology Council of Taiwan (R.O.C.). Fermilab is managed by FermiForward Discovery Group, LLC, acting under Contract No.\ 89243024CSC000002. \end{acknowledgments}

\bibliography{reference}

@article{accardi2023,
  title = {Light Quark and Antiquark Constraints from New Electroweak Data},
  author = {Accardi, A. and others},
  year = 2023,
  month = jun,
  journal = {Phys. Rev. D},
  volume = {107},
  number = {JLAB-THY-23-3782},
  eprint = {2303.11509},
  primaryclass = {hep-ph},
  pages = {113005},
  doi = {10.1103/PhysRevD.107.113005},
  url = {https://link.aps.org/doi/10.1103/PhysRevD.107.113005},
  urldate = {2023-07-14},
  archiveprefix = {arXiv}
}

@article{ackerstaff1998,
  title = {Flavor Asymmetry of the Light Quark Sea from Semi-Inclusive Deep-Inelastic Scattering},
  author = {Ackerstaff, K. and others},
  year = 1998,
  month = dec,
  journal = {Phys. Rev. Lett.},
  volume = {81},
  number = {25},
  eprint = {hep-ex/9807013},
  pages = {5519--5523},
  publisher = {American Physical Society},
  doi = {10.1103/PhysRevLett.81.5519},
  url = {https://link.aps.org/doi/10.1103/PhysRevLett.81.5519},
  archiveprefix = {arXiv},
  collaboration = {HERMES Collaboration}
}

@article{adam2021,
  title = {Measurements of {{$W$}} and {{$Z/\gamma^*$}} cross sections and their ratios in {{$p + p$}} collisions at {{RHIC}}},
  author = {Adam, J. and others},
  year = 2021,
  month = jan,
  journal = {Phys. Rev. D},
  volume = {103},
  number = {1},
  eprint = {2011.04708},
  primaryclass = {nucl-ex},
  pages = {012001},
  issn = {2470-0010, 2470-0029},
  doi = {10.1103/PhysRevD.103.012001},
  url = {https://link.aps.org/doi/10.1103/PhysRevD.103.012001},
  urldate = {2024-01-01},
  archiveprefix = {arXiv},
  collaboration = {STAR Collaboration},
  langid = {english}
}

@article{agostinelli2003,
  title = {{{GEANT4}}---a Simulation Toolkit},
  author = {Agostinelli, S. and others},
  year = 2003,
  journal = {Nucl. Instrum. Methods Phys. Res., Sect. A},
  volume = {506},
  number = {3},
  pages = {250--303},
  issn = {0168-9002},
  doi = {10.1016/S0168-9002(03)01368-8},
  url = {https://www.sciencedirect.com/science/article/pii/S0168900203013688},
  urldate = {2023-11-17},
  collaboration = {GEANT4 Collaboration}
}

@article{aidala2019,
  title = {The {{SeaQuest}} Spectrometer at {{Fermilab}}},
  author = {Aidala, C. A. and others},
  year = 2019,
  month = jun,
  journal = {Nucl. Instrum. Methods Phys. Res., Sect. A},
  volume = {930},
  eprint = {1706.09990},
  primaryclass = {physics.ins-det},
  pages = {49--63},
  issn = {0168-9002},
  doi = {10.1016/j.nima.2019.03.039},
  url = {https://www.sciencedirect.com/science/article/pii/S016890021930347X},
  urldate = {2021-07-20},
  archiveprefix = {arXiv},
  collaboration = {FNAL E906/SeaQuest Collaboration},
  langid = {english}
}

@article{alberg2022,
  title = {Pions in Proton Structure and Everywhere Else},
  author = {Alberg, Mary and Ehinger, Lucas and Miller, Gerald A.},
  year = 2022,
  month = jun,
  journal = {Phys. Rev. D},
  volume = {105},
  number = {11},
  eprint = {2108.12439},
  primaryclass = {nucl-th},
  pages = {114054},
  publisher = {American Physical Society},
  doi = {10.1103/PhysRevD.105.114054},
  url = {https://link.aps.org/doi/10.1103/PhysRevD.105.114054},
  urldate = {2022-10-20},
  archiveprefix = {arXiv}
}

@article{alekhin2023,
  title = {Impact of {SeaQuest} data on {PDF} fits at large {{$x$}}},
  author = {Alekhin, S. and others},
  year = 2023,
  month = sep,
  journal = {Eur. Phys. J. C},
  volume = {83},
  number = {9},
  eprint = {2306.01918},
  primaryclass = {hep-ph},
  pages = {829},
  issn = {1434-6052},
  doi = {10.1140/epjc/s10052-023-11999-6},
  url = {https://doi.org/10.1140/epjc/s10052-023-11999-6},
  urldate = {2023-10-04},
  archiveprefix = {arXiv},
  langid = {english}
}

@article{allison2006,
  title = {{{GEANT4}} Developments and Applications},
  author = {Allison, J. and others},
  year = 2006,
  journal = {IEEE Trans. Nucl. Sci.},
  volume = {53},
  number = {1},
  pages = {270},
  issn = {1558-1578},
  doi = {10.1109/TNS.2006.869826},
  url = {https://ieeexplore.ieee.org/document/1610988},
  urldate = {2023-11-17}
}

@article{allison2016,
  title = {Recent Developments in {{GEANT4}}},
  author = {Allison, J. and others},
  year = 2016,
  month = nov,
  journal = {Nucl. Instrum. Methods Phys. Res., Sect. A},
  volume = {835},
  pages = {186--225},
  issn = {01689002},
  doi = {10.1016/j.nima.2016.06.125},
  url = {https://linkinghub.elsevier.com/retrieve/pii/S0168900216306957},
  urldate = {2023-11-17},
  langid = {english}
}

@article{amaudruz1991,
  title = {Gottfried sum from the ratio {${\mathit{F}}_{2}^{\mathit{n}}$/${\mathit{F}}_{2}^{\mathit{p}}$}},
  author = {Amaudruz, P. and others},
  year = 1991,
  journal = {Phys. Rev. Lett.},
  volume = {66},
  number = {21},
  pages = {2712-2715},
  publisher = {American Physical Society},
  doi = {10.1103/PhysRevLett.66.2712},
  url = {https://link.aps.org/doi/10.1103/PhysRevLett.66.2712},
  urldate = {2022-02-16},
  collaboration = {New Muon collaboration}
}

@article{arneodo1994,
  title = {Reevaluation of the {{Gottfried}} Sum},
  author = {Arneodo, M. and others},
  year = 1994,
  month = jul,
  journal = {Phys. Rev. D},
  volume = {50},
  number = {1},
  pages = {R1-R3},
  publisher = {American Physical Society},
  doi = {10.1103/PhysRevD.50.R1},
  url = {https://link.aps.org/doi/10.1103/PhysRevD.50.R1},
  urldate = {2025-09-20},
  collaboration = {New Muon collaboration}
}

@article{ball2022a,
  title = {The Path to Proton Structure at 1\% Accuracy},
  author = {Ball, Richard D. and others},
  year = 2022,
  month = may,
  journal = {Eur. Phys. J. C},
  volume = {82},
  number = {Edinburgh 2021/12, Nikhef-2021-013, TIF-UNIMI-2021-11},
  eprint = {2109.02653},
  primaryclass = {hep-ph},
  pages = {428},
  issn = {1434-6052},
  doi = {10.1140/epjc/s10052-022-10328-7},
  url = {https://doi.org/10.1140/epjc/s10052-022-10328-7},
  urldate = {2023-06-23},
  archiveprefix = {arXiv},
  collaboration = {NNPDF Collaboration},
  langid = {english}
}

@article{barlow1993,
  title = {Fitting Using Finite {{Monte Carlo}} Samples},
  author = {Barlow, Roger and Beeston, Christine},
  year = 1993,
  month = oct,
  journal = {Comput. Phys. Commun.},
  volume = {77},
  number = {2},
  pages = {219--228},
  issn = {0010-4655},
  doi = {10.1016/0010-4655(93)90005-W},
  url = {https://www.sciencedirect.com/science/article/pii/001046559390005W},
  urldate = {2021-07-27},
  langid = {english}
}

@article{bourrely2002,
  title = {A Statistical Approach for Polarized Parton Distributions},
  author = {Bourrely, C. and Soffer, J. and Buccella, F.},
  year = 2002,
  journal = {Eur. Phys. J. C},
  volume = {23},
  number = {3},
  eprint = {hep-ph/0109160},
  pages = {487--501},
  issn = {1434-6052},
  doi = {10.1007/s100520100855},
  url = {https://doi.org/10.1007/s100520100855},
  urldate = {2026-01-02},
  archiveprefix = {arXiv},
  langid = {english}
}

@article{camarda2020,
  title = {{{DYTurbo}}: Fast Predictions for {{Drell}}--{{Yan}} Processes},
  author = {Camarda, Stefano and others},
  shorttitle = {{{DYTurbo}}},
  year = 2020,
  month = mar,
  journal = {Eur. Phys. J. C},
  volume = {80},
  number = {3},
  eprint = {1910.07049},
  primaryclass = {hep-ph},
  pages = {251},
  issn = {1434-6052},
  doi = {10.1140/epjc/s10052-020-7757-5},
  url = {https://doi.org/10.1140/epjc/s10052-020-7757-5},
  urldate = {2025-05-01},
  archiveprefix = {arXiv},
  langid = {english},
  note = {[Erratum: Eur.Phys.J.C 80, 440 (2020)]}
}

@article{chang2014,
  title = {Flavor Structure of the Nucleon Sea},
  author = {Chang, Wen-Chen and Peng, Jen-Chieh},
  year = 2014,
  month = nov,
  journal = {Prog. Part. Nucl. Phys.},
  volume = {79},
  eprint = {1406.1260},
  primaryclass = {hep-ph},
  pages = {95--135},
  issn = {0146-6410},
  doi = {10.1016/j.ppnp.2014.08.002},
  url = {https://www.sciencedirect.com/science/article/pii/S0146641014000568},
  urldate = {2021-07-27},
  archiveprefix = {arXiv},
  langid = {english}
}

@article{chang2022,
  title = {Parton Distributions of Light Quarks and Antiquarks in the Proton},
  author = {Chang, Lei and Gao, Fei and Roberts, Craig D.},
  year = 2022,
  month = apr,
  journal = {Phys. Lett. B},
  volume = {829},
  eprint = {2201.07870},
  primaryclass = {hep-ph},
  pages = {137078},
  issn = {0370-2693},
  doi = {10.1016/j.physletb.2022.137078},
  url = {https://www.sciencedirect.com/science/article/pii/S037026932200212X},
  urldate = {2025-12-29},
  archiveprefix = {arXiv}
}

@article{cocuzza2021,
  title = {Bayesian {{Monte Carlo}} Extraction of the Sea Asymmetry with {{SeaQuest}} and {{STAR}} Data},
  author = {Cocuzza, C. and others},
  year = 2021,
  month = oct,
  journal = {Phys. Rev. D},
  volume = {104},
  number = {7},
  eprint = {2109.00677},
  primaryclass = {hep-ph},
  pages = {074031},
  publisher = {American Physical Society},
  doi = {10.1103/PhysRevD.104.074031},
  url = {https://link.aps.org/doi/10.1103/PhysRevD.104.074031},
  archiveprefix = {arXiv},
  collaboration = {Jefferson Lab Angular Momentum (JAM) Collaboration}
}

@article{constantinou2021,
  title = {Parton Distributions and Lattice-{{QCD}} Calculations: {{Toward 3D}} Structure},
  author = {Constantinou, Martha and others},
  shorttitle = {Parton Distributions and Lattice-{{QCD}} Calculations},
  year = 2021,
  month = nov,
  journal = {Prog. Part. Nucl. Phys.},
  volume = {121},
  eprint = {2006.08636},
  primaryclass = {hep-ph},
  pages = {103908},
  issn = {0146-6410},
  doi = {10.1016/j.ppnp.2021.103908},
  url = {https://www.sciencedirect.com/science/article/pii/S0146641021000673},
  urldate = {2023-01-31},
  archiveprefix = {arXiv},
  langid = {english}
}

@phdthesis{dove2020,
  title = {Probing the Flavor Dependence of Proton's Light-Quark Sea in the {{SeaQuest}} Experiment at {{Fermilab}}},
  author = {Dove, Jason},
  year = 2020,
  month = jul,
  url = {http://hdl.handle.net/2142/109313},
  school = {University of Illinois at Urbana-Champaign}
}

@article{dove2021,
  title = {The Asymmetry of Antimatter in the Proton},
  author = {Dove, J. and others},
  year = 2021,
  month = feb,
  journal = {Nature},
  volume = {590},
  number = {7847},
  eprint = {2103.04024},
  primaryclass = {hep-ph},
  pages = {561--565},
  publisher = {Nature Publishing Group},
  issn = {1476-4687},
  doi = {10.1038/s41586-021-03282-z},
  url = {https://www.nature.com/articles/s41586-021-03282-z},
  urldate = {2021-07-20},
  archiveprefix = {arXiv},
  collaboration = {FNAL E906/SeaQuest Collaboration},
  langid = {english},
  note = {[Publisher's correction: Nature 604, E26 (2022)]}
}

@article{dove2023,
  title = {Measurement of flavor asymmetry of light-quark sea in the proton with {{Drell-Yan}} dimuon production in {$p+p$} and {$p+d$} collisions at 120 {{GeV}}},
  author = {Dove, J. and others},
  year = 2023,
  month = sep,
  journal = {Phys. Rev. C},
  volume = {108},
  number = {3},
  eprint = {2212.12160},
  primaryclass = {hep-ph},
  pages = {035202},
  doi = {10.1103/PhysRevC.108.035202},
  url = {https://link.aps.org/doi/10.1103/PhysRevC.108.035202},
  urldate = {2023-08-11},
  archiveprefix = {arXiv},
  collaboration = {FNAL E906/SeaQuest Collaboration}
}

@article{drell1970,
  title = {Massive Lepton-Pair Production in Hadron-Hadron Collisions at High Energies},
  author = {Drell, Sidney D. and Yan, Tung-Mow},
  year = 1970,
  month = aug,
  journal = {Phys. Rev. Lett.},
  volume = {25},
  number = {5},
  pages = {316--320},
  publisher = {American Physical Society},
  doi = {10.1103/PhysRevLett.25.316},
  url = {https://link.aps.org/doi/10.1103/PhysRevLett.25.316},
  urldate = {2021-07-14},
  note = {[Erratum: Phys.Rev.Lett. 25, 902 (1970)]}
}

@article{ehlers2014,
  title = {Nuclear Effects in the Proton-Deuteron {{Drell-Yan}} Process},
  author = {Ehlers, P. J. and others},
  year = 2014,
  month = jul,
  journal = {Phys. Rev. D},
  volume = {90},
  number = {1},
  eprint = {1405.2039},
  primaryclass = {hep-ph},
  pages = {014010},
  publisher = {American Physical Society},
  doi = {10.1103/PhysRevD.90.014010},
  url = {https://link.aps.org/doi/10.1103/PhysRevD.90.014010},
  urldate = {2023-10-01},
  archiveprefix = {arXiv}
}

@article{ellis1991,
  title = {Constraints on Isospin Breaking in the Light Quark Sea from the {{Drell-Yan}} Process},
  author = {Ellis, S. D. and Stirling, W. J.},
  year = 1991,
  journal = {Phys. Lett. B},
  volume = {256},
  number = {2},
  pages = {258--264},
  issn = {0370-2693},
  doi = {10.1016/0370-2693(91)90684-I},
  url = {https://www.sciencedirect.com/science/article/pii/037026939190684I},
  urldate = {2025-02-19}
}

@article{field1977,
  title = {Quark Elastic Scattering as a Source of High-Transverse-Momentum Mesons},
  author = {Field, R. D. and Feynman, R. P.},
  year = 1977,
  month = may,
  journal = {Phys. Rev. D},
  volume = {15},
  number = {9},
  pages = {2590--2616},
  publisher = {American Physical Society},
  doi = {10.1103/PhysRevD.15.2590},
  url = {https://link.aps.org/doi/10.1103/PhysRevD.15.2590},
  urldate = {2022-07-23}
}

@article{friedman1972,
  title = {Deep {{Inelastic Electron Scattering}}},
  author = {Friedman, J I and Kendall, H W},
  year = 1972,
  journal = {Annu. Rev. Nucl. Sci.},
  volume = {22},
  number = {1},
  pages = {203--254},
  doi = {10.1146/annurev.ns.22.120172.001223},
  url = {https://doi.org/10.1146/annurev.ns.22.120172.001223},
  urldate = {2022-02-05}
}

@article{garvey2001,
  title = {Flavor Asymmetry of Light Quarks in the Nucleon Sea},
  author = {Garvey, G. T. and Peng, J. -C.},
  year = 2001,
  month = jan,
  journal = {Prog. Part. Nucl. Phys.},
  volume = {47},
  number = {1},
  eprint = {nucl-ex/0109010},
  pages = {203--243},
  issn = {0146-6410},
  doi = {10.1016/S0146-6410(01)00155-7},
  url = {http://www.sciencedirect.com/science/article/pii/S0146641001001557},
  urldate = {2018-09-24},
  archiveprefix = {arXiv}
}

@article{geesaman2019,
  title = {The Sea of Quarks and Antiquarks in the Nucleon},
  author = {Geesaman, D. F. and Reimer, P. E.},
  year = 2019,
  month = mar,
  journal = {Rept. Prog. Phys.},
  volume = {82},
  number = {4},
  eprint = {1812.10372},
  primaryclass = {nucl-ex},
  pages = {046301},
  publisher = {IOP Publishing},
  issn = {0034-4885},
  doi = {10.1088/1361-6633/ab05a7},
  url = {https://dx.doi.org/10.1088/1361-6633/ab05a7},
  urldate = {2023-12-05},
  archiveprefix = {arXiv},
  langid = {english}
}

@article{gottfried1967,
  title = {Sum {{Rule}} for {{High-Energy Electron-Proton Scattering}}},
  author = {Gottfried, Kurt},
  year = 1967,
  month = jun,
  journal = {Phys. Rev. Lett.},
  volume = {18},
  number = {25},
  pages = {1174--1177},
  publisher = {American Physical Society},
  doi = {10.1103/PhysRevLett.18.1174},
  url = {https://link.aps.org/doi/10.1103/PhysRevLett.18.1174},
  urldate = {2022-02-16}
}

@misc{harland-lang2025,
  title = {A reassessment of the role of high {{$x$}} data on the {{MSHT}} global {{PDF}} fit},
  author = {Harland-Lang, L. A. and others},
  year = 2025,
  month = oct,
  eprint = {2510.03753},
  primaryclass = {hep-ph},
  doi = {10.48550/arXiv.2510.03753},
  url = {http://arxiv.org/abs/2510.03753},
  urldate = {2025-12-01},
  archiveprefix = {arXiv}
}

@article{hawker1998,
  title = {Measurement of the Light Antiquark Flavor Asymmetry in the Nucleon Sea},
  author = {Hawker, E. A. and others},
  year = 1998,
  month = apr,
  journal = {Phys. Rev. Lett.},
  volume = {80},
  number = {17},
  eprint = {hep-ex/9803011},
  pages = {3715--3718},
  publisher = {American Physical Society},
  doi = {10.1103/PhysRevLett.80.3715},
  url = {https://link.aps.org/doi/10.1103/PhysRevLett.80.3715},
  archiveprefix = {arXiv},
  collaboration = {FNAL E866/NuSea Collaboration}
}

@article{hou2018,
  title = {{{CT14}} Intrinsic Charm Parton Distribution Functions from {{CTEQ-TEA}} Global Analysis},
  author = {Hou, Tie-Jiun and others},
  year = 2018,
  month = feb,
  journal = {J. High Energ. Phys.},
  volume = {2018},
  number = {2},
  eprint = {1707.00657},
  primaryclass = {hep-ph},
  pages = {59},
  issn = {1029-8479},
  doi = {10.1007/JHEP02(2018)059},
  url = {https://doi.org/10.1007/JHEP02(2018)059},
  urldate = {2022-07-30},
  archiveprefix = {arXiv},
  langid = {english}
}

@article{kalman1960,
  title = {A {{New Approach}} to {{Linear Filtering}} and {{Prediction Problems}}},
  author = {Kalman, R. E.},
  year = 1960,
  month = mar,
  journal = {J. Basic Eng.},
  volume = {82},
  number = {1},
  pages = {35--45},
  issn = {0021-9223},
  doi = {10.1115/1.3662552},
  url = {https://asmedigitalcollection.asme.org/fluidsengineering/article/82/1/35/397706/A-New-Approach-to-Linear-Filtering-and-Prediction},
  urldate = {2023-05-22},
  langid = {english}
}

@article{kumano1998,
  title = {Flavor Asymmetry of Antiquark Distributions in the Nucleon},
  author = {Kumano, S.},
  year = 1998,
  month = sep,
  journal = {Phys. Rep.},
  volume = {303},
  number = {4},
  eprint = {hep-ph/9702367},
  pages = {183--257},
  issn = {0370-1573},
  doi = {10.1016/S0370-1573(98)00016-7},
  url = {https://www.sciencedirect.com/science/article/pii/S0370157398000167},
  urldate = {2022-07-23},
  archiveprefix = {arXiv},
  langid = {english}
}

@phdthesis{leung2024,
  title = {Probing Parton Distributions in Proton using {{Drell-Yan}} and Charmonium Production in {{$p+p$}} and {{$p+d$}} Interactions with 120 {{GeV}} Proton Beam at {{Fermilab}}},
  author = {Leung, Ching Him},
  year = 2024,
  month = mar,
  url = {https://hdl.handle.net/2142/124206},
  school = {University of Illinois at Urbana-Champaign}
}

@article{leung2024a,
  title = {Measurement of {{$J/\psi$}} and {{$\psi\left(2S\right)$}} production in {{$p+p$}} and {{$p+d$}} interactions at {{120 GeV}}},
  author = {Leung, C. H. and others},
  year = 2024,
  month = sep,
  journal = {Phys. Lett. B},
  volume = {858},
  eprint = {2406.11459},
  primaryclass = {hep-ex},
  pages = {139032},
  issn = {0370-2693},
  doi = {10.1016/j.physletb.2024.139032},
  url = {https://www.sciencedirect.com/science/article/pii/S0370269324005902},
  urldate = {2024-06-18},
  archiveprefix = {arXiv},
  collaboration = {FNAL E906/SeaQuest Collaboration}
}

@article{londergan2010,
  title = {Charge Symmetry at the Partonic Level},
  author = {Londergan, J. T. and Peng, J. C. and Thomas, A. W.},
  year = 2010,
  month = jul,
  journal = {Rev. Mod. Phys.},
  volume = {82},
  number = {3},
  eprint = {0907.2352},
  primaryclass = {hep-ph},
  pages = {2009--2052},
  publisher = {American Physical Society},
  doi = {10.1103/RevModPhys.82.2009},
  url = {https://link.aps.org/doi/10.1103/RevModPhys.82.2009},
  urldate = {2021-07-14},
  archiveprefix = {arXiv}
}

@article{NA51:1994xrz,
  title = {Study of the Isospin Symmetry Breaking in the Light Quark Sea of the Nucleon from the {{Drell-Yan}} Process},
  author = {Baldit, A. and others},
  year = 1994,
  journal = {Phys. Lett. B},
  volume = {332},
  number = {1},
  pages = {244--250},
  issn = {0370-2693},
  doi = {10.1016/0370-2693(94)90884-2},
  url = {https://www.sciencedirect.com/science/article/pii/0370269394908842},
  urldate = {2024-08-19},
  collaboration = {NA51}
}

@article{pate2023,
  title = {Estimation of Combinatoric Background in Seaquest Using an Event-Mixing Method},
  author = {Pate, S. F. and others},
  year = 2023,
  journal = {JINST},
  volume = {18},
  number = {10},
  eprint = {2302.04152},
  primaryclass = {hep-ex},
  pages = {P10032},
  publisher = {IOP Publishing},
  issn = {1748-0221},
  doi = {10.1088/1748-0221/18/10/P10032},
  url = {https://dx.doi.org/10.1088/1748-0221/18/10/P10032},
  urldate = {2024-01-21},
  archiveprefix = {arXiv},
  collaboration = {FNAL E906/SeaQuest Collaboration},
  langid = {english}
}

@article{peng1998,
  title = {{{$\bar{d}/\bar{u}$}} asymmetry and the origin of the nucleon sea},
  author = {Peng, J. C. and others},
  year = 1998,
  month = sep,
  journal = {Phys. Rev. D},
  volume = {58},
  number = {9},
  eprint = {hep-ph/9804288},
  pages = {092004},
  publisher = {American Physical Society},
  doi = {10.1103/PhysRevD.58.092004},
  url = {https://link.aps.org/doi/10.1103/PhysRevD.58.092004},
  archiveprefix = {arXiv},
  collaboration = {FNAL E866/NuSea Collaboration}
}

@phdthesis{prasad2020,
  title = {Measurement of high-mass dimuon production for {{$p+p$}} and {{$p+d$}} collisions with 120 {{GeV}} proton beam at {{Fermilab}}},
  author = {Prasad, Shivangi},
  year = 2020,
  month = nov,
  url = {http://hdl.handle.net/2142/109365},
  school = {University of Illinois at Urbana-Champaign}
}

@article{soffer2019,
  title = {On the Flavor Structure of the Light-Quark Sea Distributions},
  author = {Soffer, Jacques and Bourrely, Claude},
  year = 2019,
  month = nov,
  journal = {Nucl. Phys. A},
  volume = {991},
  pages = {121607},
  issn = {03759474},
  doi = {10.1016/j.nuclphysa.2019.08.001},
  url = {https://linkinghub.elsevier.com/retrieve/pii/S0375947419301757},
  urldate = {2023-03-02},
  langid = {english}
}

@incollection{speth1997,
  title = {Mesonic {{Contributions}} to the {{Spin}} and {{Flavor Structure}} of the {{Nucleon}}},
  author = {Speth, J. and Thomas, A. W.},
  booktitle = {Adv. {{Nucl}}. {{Phys}}.},
  year = 1997,
  volume = {24},
  pages = {83--149},
  publisher = {Springer US},
  address = {Boston, MA},
  doi = {10.1007/0-306-47073-X_2},
  url = {https://doi.org/10.1007/0-306-47073-X_2},
  urldate = {2026-01-02},
  isbn = {978-0-306-47073-8},
  langid = {english}
}

@article{stein1975,
  title = {Electron scattering at {{4\ifmmode^\circ\else\textdegree\fi{}}} with energies of {{4.5-20 GeV}}},
  author = {Stein, S. and others},
  year = 1975,
  journal = {Phys. Rev. D},
  volume = {12},
  number = {7},
  pages = {1884},
  publisher = {American Physical Society},
  doi = {10.1103/PhysRevD.12.1884},
  url = {https://link.aps.org/doi/10.1103/PhysRevD.12.1884},
  urldate = {2025-02-13}
}

@article{szczurek1996,
  title = {On the Flavour Structure of the Constituent Quark},
  author = {Szczurek, Antoni and Buchmann, Alfons J. and Faessler, Amand},
  year = 1996,
  journal = {J. Phys. G},
  volume = {22},
  number = {12},
  eprint = {nucl-th/9609042},
  pages = {1741--1750},
  issn = {0954-3899},
  doi = {10.1088/0954-3899/22/12/004},
  url = {https://doi.org/10.1088/0954-3899/22/12/004},
  urldate = {2026-01-02},
  archiveprefix = {arXiv},
  langid = {english}
}

@article{thomas1983,
  title = {A Limit on the Pionic Component of the Nucleon through {{SU}}(3) Flavour Breaking in the Sea},
  author = {Thomas, A. W.},
  year = 1983,
  journal = {Phys. Lett. B},
  volume = {126},
  number = {1},
  pages = {97--100},
  issn = {0370-2693},
  doi = {10.1016/0370-2693(83)90026-6},
  url = {https://www.sciencedirect.com/science/article/pii/0370269383900266},
  urldate = {2026-01-02}
}

@article{towell2001,
  title = {Improved measurement of the {{$\bar{d}/\bar{u}$}} asymmetry in the nucleon sea},
  author = {Towell, R. S. and others},
  year = 2001,
  month = aug,
  journal = {Phys. Rev. D},
  volume = {64},
  number = {5},
  eprint = {hep-ex/0103030},
  pages = {052002},
  publisher = {American Physical Society},
  doi = {10.1103/PhysRevD.64.052002},
  url = {https://link.aps.org/doi/10.1103/PhysRevD.64.052002},
  archiveprefix = {arXiv},
  collaboration = {FNAL E866/NuSea Collaboration}
}

@article{vogt2000a,
  title = {Physics of the Nucleon Sea Quark Distributions},
  author = {Vogt, R.},
  year = 2000,
  journal = {Prog. Part. Nucl. Phys.},
  volume = {45},
  eprint = {hep-ph/0011298},
  pages = {S105-S169},
  issn = {0146-6410},
  doi = {10.1016/S0146-6410(00)90012-7},
  url = {https://www.sciencedirect.com/science/article/pii/S0146641000900127},
  urldate = {2025-02-26},
  archiveprefix = {arXiv}
}

@article{yin2023,
  title = {All-{{Orders Evolution}} of {{Parton Distributions}}: {{Principle}}, {{Practice}}, and {{Predictions}}},
  author = {Yin, Pei-Lin and others},
  shorttitle = {All-{{Orders Evolution}} of {{Parton Distributions}}},
  year = 2023,
  journal = {Chin. Phys. Lett.},
  volume = {40},
  number = {9},
  eprint = {2306.03274},
  primaryclass = {hep-ph},
  pages = {091201},
  publisher = {{Chinese Physical Society and IOP Publishing Ltd}},
  issn = {0256-307X},
  doi = {10.1088/0256-307X/40/9/091201},
  url = {https://doi.org/10.1088/0256-307X/40/9/091201},
  urldate = {2025-12-29},
  archiveprefix = {arXiv},
  langid = {english}
}

\clearpage
\onecolumngrid
\appendix
\section{End Matter}
\begin{table*}[htbp!]
	\centering
	\caption{The acceptance of the spectrometer for different $x_1$ and $x_2$ bins.
		In each cell, the first value is the acceptance,
		the second is the average $x_1$, the third is the average $x_2$,
		and the fourth is the average mass in \unit{\GeV/c^2}.}
	\label{tab:acceptance}
	\begin{adjustbox}{max width=\linewidth}
		\begin{tabular}{c|c|c|c|c|c|c|c|c|c|c}
			\hline
			\hline
			\diagbox{$x_2$}{$x_1$} & $0.30$--$0.35$                                                           & $0.35$--$0.40$                                                           & $0.40$--$0.45$                                                           & $0.45$--$0.50$                                                           & $0.50$--$0.55$                                                           & $0.55$--$0.60$                                                           & $0.60$--$0.65$                                                           & $0.65$--$0.70$                                                           & $0.70$--$0.75$                                                           & $0.75$--$0.80$                                                           \\
			\hline
			$0.130$--$0.160$       &                                                                          &                                                                          &                                                                          &                                                                          &                                                                          & \begin{tabular}{@{}c@{}} $1.19\%$\\$0.590$\\$0.157$\\$4.54$\end{tabular} & \begin{tabular}{@{}c@{}} $2.47\%$\\$0.628$\\$0.153$\\$4.60$\end{tabular} & \begin{tabular}{@{}c@{}} $3.22\%$\\$0.676$\\$0.148$\\$4.67$\end{tabular} & \begin{tabular}{@{}c@{}} $3.81\%$\\$0.723$\\$0.144$\\$4.77$\end{tabular} & \begin{tabular}{@{}c@{}} $4.35\%$\\$0.772$\\$0.143$\\$4.91$\end{tabular} \\
			\hline
			$0.160$--$0.195$       &                                                                          &                                                                          &                                                                          & \begin{tabular}{@{}c@{}} $1.03\%$\\$0.489$\\$0.191$\\$4.55$\end{tabular} & \begin{tabular}{@{}c@{}} $1.97\%$\\$0.529$\\$0.184$\\$4.63$\end{tabular} & \begin{tabular}{@{}c@{}} $2.66\%$\\$0.575$\\$0.178$\\$4.73$\end{tabular} & \begin{tabular}{@{}c@{}} $3.64\%$\\$0.623$\\$0.176$\\$4.89$\end{tabular} & \begin{tabular}{@{}c@{}} $4.08\%$\\$0.673$\\$0.176$\\$5.08$\end{tabular} & \begin{tabular}{@{}c@{}} $4.87\%$\\$0.722$\\$0.176$\\$5.27$\end{tabular} & \begin{tabular}{@{}c@{}} $5.40\%$\\$0.771$\\$0.175$\\$5.44$\end{tabular} \\
			\hline
			$0.195$--$0.240$       &                                                                          & \begin{tabular}{@{}c@{}} $0.04\%$\\$0.393$\\$0.235$\\$4.54$\end{tabular} & \begin{tabular}{@{}c@{}} $0.66\%$\\$0.433$\\$0.226$\\$4.64$\end{tabular} & \begin{tabular}{@{}c@{}} $1.51\%$\\$0.476$\\$0.218$\\$4.77$\end{tabular} & \begin{tabular}{@{}c@{}} $2.50\%$\\$0.524$\\$0.215$\\$4.97$\end{tabular} & \begin{tabular}{@{}c@{}} $3.34\%$\\$0.574$\\$0.215$\\$5.20$\end{tabular} & \begin{tabular}{@{}c@{}} $4.32\%$\\$0.623$\\$0.215$\\$5.42$\end{tabular} & \begin{tabular}{@{}c@{}} $5.08\%$\\$0.673$\\$0.215$\\$5.64$\end{tabular} & \begin{tabular}{@{}c@{}} $5.65\%$\\$0.723$\\$0.215$\\$5.85$\end{tabular} & \begin{tabular}{@{}c@{}} $6.06\%$\\$0.771$\\$0.214$\\$6.02$\end{tabular} \\
			\hline
			$0.240$--$0.290$       & \begin{tabular}{@{}c@{}} $0.03\%$\\$0.343$\\$0.279$\\$4.63$\end{tabular} & \begin{tabular}{@{}c@{}} $0.26\%$\\$0.383$\\$0.267$\\$4.76$\end{tabular} & \begin{tabular}{@{}c@{}} $0.95\%$\\$0.427$\\$0.264$\\$4.97$\end{tabular} & \begin{tabular}{@{}c@{}} $2.02\%$\\$0.474$\\$0.263$\\$5.23$\end{tabular} & \begin{tabular}{@{}c@{}} $3.00\%$\\$0.524$\\$0.261$\\$5.49$\end{tabular} & \begin{tabular}{@{}c@{}} $4.03\%$\\$0.574$\\$0.262$\\$5.76$\end{tabular} & \begin{tabular}{@{}c@{}} $4.92\%$\\$0.623$\\$0.262$\\$6.00$\end{tabular} & \begin{tabular}{@{}c@{}} $5.56\%$\\$0.673$\\$0.262$\\$6.23$\end{tabular} & \begin{tabular}{@{}c@{}} $6.11\%$\\$0.722$\\$0.262$\\$6.46$\end{tabular} & \begin{tabular}{@{}c@{}} $6.32\%$\\$0.771$\\$0.262$\\$6.68$\end{tabular} \\
			\hline
			$0.290$--$0.350$       & \begin{tabular}{@{}c@{}} $0.04\%$\\$0.338$\\$0.323$\\$4.92$\end{tabular} & \begin{tabular}{@{}c@{}} $0.45\%$\\$0.379$\\$0.318$\\$5.16$\end{tabular} & \begin{tabular}{@{}c@{}} $1.38\%$\\$0.425$\\$0.316$\\$5.44$\end{tabular} & \begin{tabular}{@{}c@{}} $2.46\%$\\$0.474$\\$0.315$\\$5.75$\end{tabular} & \begin{tabular}{@{}c@{}} $3.40\%$\\$0.524$\\$0.316$\\$6.04$\end{tabular} & \begin{tabular}{@{}c@{}} $4.18\%$\\$0.573$\\$0.315$\\$6.32$\end{tabular} & \begin{tabular}{@{}c@{}} $5.02\%$\\$0.624$\\$0.314$\\$6.59$\end{tabular} & \begin{tabular}{@{}c@{}} $5.50\%$\\$0.673$\\$0.314$\\$6.84$\end{tabular} & \begin{tabular}{@{}c@{}} $6.19\%$\\$0.722$\\$0.313$\\$7.09$\end{tabular} & \begin{tabular}{@{}c@{}} $6.10\%$\\$0.772$\\$0.315$\\$7.36$\end{tabular} \\
			\hline
			$0.350$--$0.450$       & \begin{tabular}{@{}c@{}} $0.08\%$\\$0.337$\\$0.385$\\$5.38$\end{tabular} & \begin{tabular}{@{}c@{}} $0.64\%$\\$0.377$\\$0.387$\\$5.69$\end{tabular} & \begin{tabular}{@{}c@{}} $1.48\%$\\$0.425$\\$0.386$\\$6.03$\end{tabular} & \begin{tabular}{@{}c@{}} $2.32\%$\\$0.475$\\$0.385$\\$6.38$\end{tabular} & \begin{tabular}{@{}c@{}} $3.09\%$\\$0.523$\\$0.385$\\$6.69$\end{tabular} & \begin{tabular}{@{}c@{}} $3.80\%$\\$0.573$\\$0.384$\\$6.99$\end{tabular} & \begin{tabular}{@{}c@{}} $4.37\%$\\$0.624$\\$0.383$\\$7.29$\end{tabular} & \begin{tabular}{@{}c@{}} $4.68\%$\\$0.671$\\$0.379$\\$7.53$\end{tabular} & \begin{tabular}{@{}c@{}} $4.93\%$\\$0.721$\\$0.380$\\$7.81$\end{tabular} & \begin{tabular}{@{}c@{}} $4.85\%$\\$0.772$\\$0.378$\\$8.06$\end{tabular} \\
			\hline
			\hline
		\end{tabular}
	\end{adjustbox}
\end{table*}
 \begin{figure*}[htbp!]
	\begin{subfigure}{0.45\linewidth}
		\includegraphics[width=\linewidth]{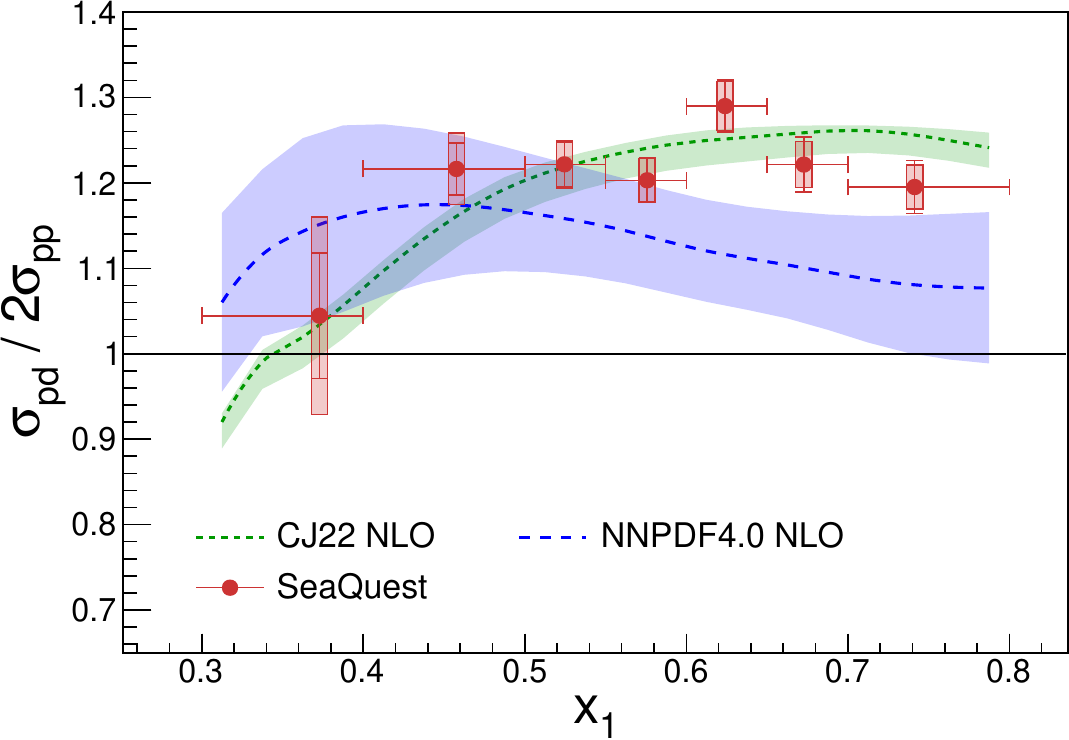}
	\end{subfigure}
	\begin{subfigure}{0.45\linewidth}
		\includegraphics[width=\linewidth]{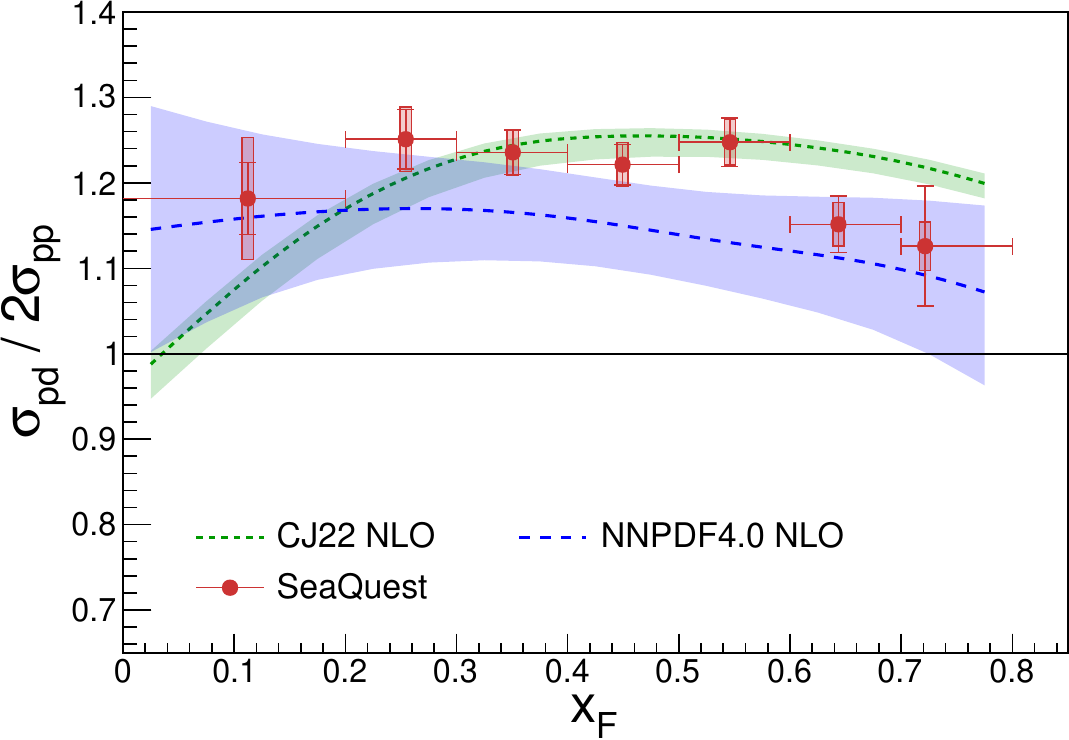}
	\end{subfigure}
	\caption{Measured $\sigma_{pd}/2\sigma_{pp}$ Drell-Yan cross section ratio as functions of $x_1$ (left), and $x_F$ (right)
		from SeaQuest compared calculations using
		CJ22~\cite{accardi2023}, and NNPDF4.0~\cite{ball2022a}.
		The error bands on the theoretical calculations correspond to the \SI{68}{\percent} confidence level from the PDFs.}
	\label{fig:csr_x1_xF}
\end{figure*}

\begin{table*}[htbp!]
	\begin{center}
		\caption{The measured $\sigma_{pd}/2\sigma_{pp}$ cross section ratio as functions of $x_1$ and $x_F$.
			The first uncertainty is statistical and the second systematic.}
		\label{tab:csr_x1_xF}
		\begin{adjustbox}{max width=\linewidth}
			{
\renewcommand{\arraystretch}{1.5}
\begin{tabular}{ccccc|ccccc}
	\hline\hline
	$x_{1}$ range    & $\expval{x_{1}}$ & \begin{tabular}{@{}c@{}} $\expval{p_{T}}$\\(\unit{\GeV/c})\end{tabular} & \begin{tabular}{@{}c@{}}$\expval{M}$\\(\unit{\GeV/c^2})\end{tabular} &$\sigma_{pd}/2\sigma_{pp}$ &   $x_{F}$ range    & $\expval{x_{F}}$ & \begin{tabular}{@{}c@{}} $\expval{p_{T}}$\\(\unit{\GeV/c})\end{tabular} & \begin{tabular}{@{}c@{}}$\expval{M}$\\(\unit{\GeV/c^2})\end{tabular} & $\sigma_{pd}/2\sigma_{pp}$                  \\ \hline
	$0.30$--$0.40$ & $0.373$          & $0.697$ & $5.16$ & $1.045\pm0.074\pm0.116$     & $-0.10$--$0.20$     & $0.112$          & $0.771$ & $5.50$ & $1.182\pm0.042\pm0.071$\\
	$0.40$--$0.50$ & $0.458$          & $0.759$ & $5.19$ & $1.216\pm0.030\pm0.042$     & $0.20$--$0.30$      & $0.254$          & $0.779$ & $5.27$ & $1.251\pm0.035\pm0.038$\\
	$0.50$--$0.55$ & $0.525$          & $0.765$ & $5.21$ & $1.222\pm0.028\pm0.026$     & $0.30$--$0.40$      & $0.351$          & $0.777$ & $5.20$ & $1.236\pm0.026\pm0.027$\\
	$0.55$--$0.60$ & $0.576$          & $0.764$ & $5.24$ & $1.203\pm0.026\pm0.025$     & $0.40$--$0.50$      & $0.449$          & $0.772$ & $5.18$ & $1.221\pm0.024\pm0.026$\\
	$0.60$--$0.65$ & $0.624$          & $0.764$ & $5.28$ & $1.290\pm0.030\pm0.029$     & $0.50$--$0.60$      & $0.546$          & $0.766$ & $5.17$ & $1.248\pm0.028\pm0.027$\\
	$0.65$--$0.70$ & $0.673$          & $0.763$ & $5.32$ & $1.221\pm0.032\pm0.027$     & $0.60$--$0.70$      & $0.644$          & $0.763$ & $5.15$ & $1.151\pm0.033\pm0.025$\\ 
	$0.70$--$0.80$ & $0.741$          & $0.761$ & $5.36$ & $1.195\pm0.031\pm0.026$     & $0.70$--$0.80$      & $0.722$          & $0.761$ & $4.95$ & $1.126\pm0.070\pm0.028$\\ \hline\hline
\end{tabular}
}
 		\end{adjustbox}
	\end{center}
\end{table*}

\end{document}